\documentstyle[11pt,epsf]{article}

\begin{document}

\begin{flushright}
SU-4240-638 \\
March 1997
\end{flushright}

\begin{center}
\vspace{24pt}

{\Large \bf The Hausdorff Dimension of Surfaces in Two-Dimensional
 Quantum Gravity Coupled to Ising Minimal Matter}
\vspace{24pt}

{\large \sl Mark~J.~Bowick, Varghese~John and 
 Gudmar~Thorleifsson}  \\
\vspace{6pt}
Physics Department, Syracuse University \\
Syracuse, NY 13244-1130, USA

\vspace{15pt}

\begin{abstract}
Within the framework of string field theory the intrinsic Hausdorff
dimension $d_H$ of the ensemble of surfaces in two-dimensional quantum
gravity has recently been claimed to be $2m$ for the class
of unitary minimal models $(p=m+1,q=m)$.  This contradicts recent
results from numerical simulations, which consistently find $d_H
\approx 4$ in the cases $m=2$, 3, 5 and $\infty$.  The string field
calculations rely on identifying the scaling behavior of geodesic
distance and area with respect to a common length scale $l$. 
This length scale is introduced by formulating the 
models on a disk with fixed boundary length $l$. 
In this paper we study the relationship between the mean area and the 
boundary length for pure gravity and the Ising model coupled to gravity. 
We discuss how this relationship is modified by relevant perturbations 
in the Ising model.  We discuss how this leads to a modified value for the 
Hausdorff dimension.
 
\end{abstract}
\end{center}

\vspace{20pt}

\section{Introduction}

An important characterization of surfaces in quantum
gravity is the {\it fractal} or Hausdorff dimension $d_H$.
The Hausdorff dimension governs the power-law relation
between two reparametrization-invariant quantities with
dimensions of volume and length;  $V \propto r^{d_H}$.
To introduce a quantity with the dimension of length it is convenient
to use the notion of geodesic distance between two points
on the surface.  In two dimensions the above relation tells
us how the area within geodesic distance $r$ from a marked point
scales with $r$.  It is important to realize that since we integrate
over all metrics in quantum gravity, the above notion of area has to 
be replaced by its average value. 

Since a lattice possess a fundamental length scale, it is 
natural to use the above relation to measure the Hausdorff dimension
in numerical simulations.  This was done recently 
\cite{num1,num2,num3}, using
finite size scaling, for the case of pure two dimensional
gravity, and for matter fields (Ising and 3-state Potts model)
coupled to gravity.  For these models the Hausdorff dimension
appears to be universal and have the value 4.
It is surprising that there is no apparent evidence of the
back-reaction of matter on the Hausdorff dimension; in contrast
the string susceptibility exponent is known to
depend strongly on the central charge of the matter system.  
This universality does not seem to be valid for
matter systems with central charge $c > 1$ coupled to gravity 
{---} for $c$ sufficiently large $d_H$ approaches 2 \cite{num2}.

The analytical predictions for the Hausdorff
dimension, on the other hand, are wide and varied 
\cite{kawni,kawamoto,iikmns,kost2}.  
In the continuum formulation
there is no natural notion of a length scale {---} it has to be
introduced by hand.  Both the concept of geodesic distance
and area are then defined relative to this length scale.
All of this may be accomplished by formulating the model on
a manifold with a boundary of fixed length, such as the disk.
Using either the continuum Liouville or the matrix model formulation
one can determine the area scaling behavior of the disk amplitude 
with boundary length $l$ \cite{moore1,goulian1}. 
Then either using a transfer
matrix formulation \cite{kkmw,wat1}, or a string field theory formulation
\cite{iikmns}, 
one can calculate the scaling behavior of the boundary loop 
length with respect to an evolution parameter.  It can be argued 
that the evolution parameter defines a notion of
geodesic distance on the surface. 
These two relations in principle determine 
the  Hausdorff dimension.  
In the case of pure gravity, where a complete transfer matrix
formalism exists, one finds Hausdorff dimension 4.
Recently a string field theory 
describing the $c < 1$ $m$-th minimal 
model coupled to gravity has been formulated 
\cite{iikmns,ishkaw,jero,sftmat,fikn}.
This formulation, combined with the assumption of 
``classical'' area scaling        
$A \sim l^2$, leads to a prediction $d_H = 2m$. 
This result has also been obtained in the loop 
gas formulation of the 
string field theory Hamiltonian \cite{kost2}.

In this paper we will argue that the scaling of area with
boundary length depends crucially on the approach to the
continuum limit in the case of matter coupled to gravity.
We show in the case of the Ising model how the perturbation by the thermal 
operator changes the dependence of the mean area on the boundary length, and 
this in turn changes the Hausdorff dimension from $d_H =6$ to $d_H =4$.
This phenomena is caused by the subtle interplay between the manner in which 
the matter and gravitational correlation lengths diverge in the 
thermodynamic limit.


In Section 2 we derive the scaling of geodesic distance
with boundary length, which follows from the string field
theory Hamiltonian for $c<1$ non{--}critical strings.
In Section 3 we use matrix model techniques to compute the
disk amplitude and the resulting scaling of the mean area
with boundary length for pure gravity and the Ising model coupled to gravity.
In Section 4 we discuss the derivation of the Hausdorff dimension
for these two models.
Finally Section 5 summarizes our results.

\section{String field theory}

In this section we provide a brief review of the 
non-critical string theory for $c<1$ matter \cite{iikmns,ishkaw,sftmat}.  
This will enable us to extract the scaling
of geodesic distance with boundary length.  For simplicity we
will describe the case $c=0$. 

An essential feature of both the transfer matrix formulation, and
the string field theory, is the choice of a particular ADM or temporal
type gauge, in which time corresponds to geodesic distance $r$. 
The Hamiltonian which evolves a loop of length $l$
(corresponding to a spatial hyper-surface) is assumed to be
\begin{equation}
\label{*21}
{\cal H}_{disk} \;=\; \int_{0}^{\infty}\!\!dl_1 \int_{0}^{\infty}\!\! dl_2 
\,\psi^{\dagger}(\l_1) \,\psi^{\dagger}(l_2) \,\psi (l_1+l_2)
\,(l_1+l_2)  \:+\:  \int_{0}^{\infty}\!\! dl \,\rho(l) \,\psi(l)\,,
\end{equation}
where $\psi^{\dagger}(l)$ and $\psi(l)$ are creation and annihilation
operators for a loop of length $l$, which satisfy 
canonical commutation relations 
$[\psi(l),\psi^{\dagger}(l^{\prime})] = \delta(l -l^{\prime})$.
The first term corresponds to a cubic string vertex and the second
term to the annihilation of a single string (a tadpole term).
The absence of a kinetic term follows from a few simple assumptions
\cite{iikmns}.
The disk partition function $w(l)$ is defined by
\begin{equation}
\label{*22}
w(l,\mu) \;=\; \lim_{r \rightarrow \infty} \left <0 \left | 
{\rm e}^{\textstyle -r {\cal H}_{disk}} \right | 0 \right >,
\end{equation}
where $\mu$ is the cosmological constant.
The kernel of the tadpole term $\rho(l)$ is determined by matching
the disk amplitude, computed from the above Hamiltonian, with
the result from matrix model calculations.  One can show
subsequently that the resulting Hamiltonian satisfies the
correct Schwinger{--}Dyson equations. 

This formalism has been generalized to include all the  
$(p,q)${--}models.  The operators $\psi^{\dagger}(l)$
and $\psi(l)$ now correspond to the creation and annihilation
of a loop of length $l$ with a {\it fixed} uniform matter 
configuration.  The cubic vertex term in the Hamiltonian
Eq.~(\ref{*21}) must be generalized to include splitting of
a loop into two loops with different matter configurations.

From Eq.~(\ref{*21}) it follows by dimensional analysis that
\begin{equation}
\label{*23}
{\cal H}_{disk} \;\sim\; l^{[\psi^{\dagger}(l)] + 2},
\end{equation} 
where $[\psi^{\dagger}(l)]$ denotes the boundary length scaling
dimension of the loop creation operator.  From Eq.~(\ref{*22}) this
equals the scaling dimension of the disk amplitude,\footnote{The
relation between $\psi^{\dagger}(l)$ and $w(l)$ is even clearer if one
considers the effective Hamiltonian ${\cal H}^{\prime}$ for processes in
which one tracks only a single loop.  In \cite{ishkaw} it was shown
that
\begin{equation}
{\cal H}^{\prime}_{disk} \;=\: 2 
\int_{0}^{\infty}\!\! dl_1 \int_{0}^{\infty}\!\! dl_2
\,\psi^{\dagger}(\l_1) \,w(l_2) \,\psi (l_1+l_2)
\,(l_1+l_2) . 
\end{equation}
}
which, for $(p,q)$ conformal matter, is well known to be \cite{kpz}:
\begin{equation}
\label{*24}
[\psi^{\dagger}(l)] \;=\; [w(l)]\;=\; - \left( \frac{p+q}{q} \right) \;.
\end{equation}
Combining the last two equations we find 
\begin{equation}
\label{*25}
r \;\sim\; l^{(p-q)/q},
\end{equation}
which, in the case of the $m$-th unitary minimal model yields
\begin{equation}
\label{*26}
r \;\sim\; l^{1/m}.
\end{equation}

A different derivation of this result has been given in a loop
gas formulation, generalized to include open strings \cite{kost2}.

\section{Scaling limits of unitary minimal models}

In this section we want to relate the
mean area $\bar{A}$ of a disk to its boundary length $l$.
We will do this by using the matrix model formulation of
the disk amplitude.
We will show in Section 3.2 that this relation 
depends on how the continuum limit is taken.  Combined
with the result of Section 2, $l \sim r^m$, this gives the Hausdorff
dimension $d_H$.  

A general $(p,q)${--}model, with central charge $c=1 - 6(p-q)^2/pq$,
coupled to gravity can be defined
in terms of two differential operators $P$ and $Q$,
of degrees $p$ and $q$ respectively,
satisfying the string equation $[P,Q] = 1$ 
(for a comprehensive review see \cite{gins2}).
To obtain a surface with boundary length $b$, on the lattice,
we need to compute\footnote{Using 
this definition we
obtain a disk with fixed boundary condition for the matter
fields.  For generalizations to arbitrary boundary
conditions see \cite{taylor}.  These formulations,
on the other hand, do not allow us to study perturbations 
of the models, as we do in Section 3.2.}
$w(b) = \langle {\rm tr} \,\phi^b \rangle$.
In the continuum limit, in the spherical approximation,
this becomes 
\begin{equation}
\label{*31}w(l,\mu) \;=\; \int_{\mu}^{\infty}\!\! dx \; \left < x
 \left | {\rm e}^{\textstyle +lQ} \right | x \right > ,
\end{equation}
where $\mu$ is a cut-off identified with the cosmological constant
and $b$ has been taken to infinity so as to obtain a finite
boundary length $l$.

To evaluate the integral in Eq.~(\ref{*31}) we need, for
a particular model, both the explicit form of $Q$ 
and the string equation.  For the pure gravity and the Ising  model,
which we consider in this paper,
the operators $P$ and $Q$, together with the string equation,
may be found in \cite{gins2,difran1}.
We will calculate $w(l,\mu)$ for these two
models in the following subsections.  This will
gives us the desired scaling relations $\bar{A} \sim l^{\beta}$.

\subsection{Pure gravity ($m=2$)}

The continuum limit of the one-matrix model, at its
$k$-th multi-critical point, describes  a
$(2k-1,2)${--}matter system coupled to gravity.
In particular for $k=2$ we obtain pure gravity ($c=0$).
In this case $Q = {\rm d}^2 - u(x)$ is the Schr\"{o}dinger operator
and the potential $u(x)$ is the specific heat.
The string equation, in the
planar limit, gives $u(x) = \sqrt{x}$.  Inserting a complete
set of eigenstates for the momentum operator $p$, Eq.~(\ref{*31})
becomes 
\begin{equation}
\label{*32}
w(l,\mu) \:=\; \int_{-\infty}^{+\infty}\!\! dp \int_{\mu}^{\infty}\!\! dx
          \; {\rm e}^{\textstyle -lp^2 -\sqrt{x}\,l}  
     \;=\; \frac{2\sqrt{\pi}}{l^{5/2}} \; 
          {\rm e}^{\textstyle -\sqrt{\mu}\,l}
          \; \left[ 1 + \sqrt{\mu}\,l \right],
\end{equation}
and the mean area is        
\begin{equation}
\label{*33}
\bar{A} \;=\; - \frac{d}{d\mu} \log w(l,\mu)
        \;=\; \frac{1}{2} \; \frac{l^2}{1+\sqrt{\mu}\,l} \;.
\end{equation}

To take the thermodynamic limit we must tune $l$ and $\mu$ in such
a way that the average area $\bar{A}$ diverges.  Introducing a
dimensionless scaling variable $z = \sqrt{\mu}\,l$, we can
do this by taking $l$ to infinity in 
{\it three} different ways: $z \rightarrow 0$, 
$z = {\it const}$, and $z \rightarrow \infty$, respectively.  
For the first two cases 
Eq.(~\ref{*33}) gives 
\begin{equation}
\label{*34}
\bar{A} \;\sim\; l^2.
\end{equation}
In these two cases the boundary length diverges sufficiently
slowly that, 
in the thermodynamic limit, a vanishing fraction of the disk is on  
the boundary.   
Combining this relation with Eq.~(\ref{*26}) gives 
Hausdorff dimension $d_H = 4$.  This agrees with
other analytical calculations \cite{kawamoto,kkmw,jan5}, 
and numerical simulations \cite{num1,num2}, for pure gravity.

The third limit, $z \rightarrow \infty$, 
gives $\bar{A} \sim l$ implying $d_H = 2$.  
This, on the other hand, corresponds
in the thermodynamic limit to a disk with a finite fraction
of the area close to the boundary, akin to a Bethe lattice.
This is the branched polymer phase of the model.

\subsection{The Ising model ($m=3$)}

The Ising model, or the $(4,3)${--}model, was solved as a
two-matrix model in \cite{kaz1}, and the disk amplitude
with fixed boundary condition was calculated in       
\cite{goulian1,kost1,moore1}.  
Goulian \cite{goulian1}, in fact, calculated the disk amplitude for the
Ising model off criticality.
Since the thermal perturbation plays a crucial role
in our subsequent analysis, we will review  
this calculation here.

In the planar limit, with vanishing external magnetic field,
the differential operators $P$ and $Q$, for the Ising model
in the high-temperature phase, are
\begin{eqnarray}
\label{*345}
Q &= &\left ( {\rm d}^2 - u \right )^{3/2}_+  
   \;=\; {\rm d}^3 - {\textstyle \frac{3}{2}}u{\rm d}\,, \\ \nonumber
P &= &Q^{4/3}_+  - t\,Q^{2/3}_+ \;=\; {\rm d}^4 - 2u{\rm d}^2
+ {\textstyle \frac{1}{2}}u^2 - 
t\left ({\rm d}^2 - u\right )\,,
\end{eqnarray}
where $t$ is the deviation from the critical temperature. 
The corresponding string equation is  
\begin{equation}
\label{*35}
x \;=\; {\textstyle \frac{1}{2}} u^3 + 
{\textstyle \frac{3}{4}} tu^2 \,.
\end{equation} 
The disk amplitude Eq.~(\ref{*31}) becomes
\begin{eqnarray}
\label{*36}
w(l,\mu) &= & \int_{\mu}^{\infty} \!\!dx \; \int_{-\infty}^{+\infty}
          \!\! dp \; {\rm e}^{\textstyle -il \left (p^3 + 
          {\textstyle \frac{3}{2}} u(x,t)p \right )}  \\
     &= & \sqrt{{\textstyle \frac{2}{3}}} \;
          \int_{\mu}^{\infty} \!\! dx \; \sqrt{u(x,t)} \;
          K_{1/3}\left ( {\textstyle \frac{1}{\sqrt{2}}} 
          u^{3/2}(x,t)\,l \right ) \,,
\end{eqnarray}
where $K_\nu(y)$ is a modified Bessel function of the second kind
\cite[Eq.~8.433]{grads}.  Using the string equation Eq.~(\ref{*35})
and \cite[Eq.~5.52]{grads} we get 
\begin{equation}
\label{*37}
w(l,\mu) \;=\; \frac{2}{\sqrt{3}} \, 
      \frac{u(\mu,t)}{l} \,
     \left [ u(\mu,t) K_{4/3} (z) + t K_{2/3}(z) \right ], 
\end{equation}
where we have introduced the scaling variable 
$z = u^{3/2}(\mu,t)l/\sqrt{2}$.  The mean area is then
\begin{equation}
\label{*38}
\bar{A} \;=\; \frac{l^2 \; K_{1/3}(z)}
  {2\,z K_{4/3}(z) \; + \; 
    2^{2/3}\,t\,l^{2/3}\,z^{1/3} K_{2/3}(z) } \;.
\end{equation}

Again we consider the physical
limits $z \rightarrow 0$ and $z = {\it const}$.  We must now
specify how the critical point $t=0$ should be approached in
the continuum limit. In order to obtain scaling characteristic
of the bulk, one should approach the critical point with
the matter correlation length $\xi_M$
strictly less than the system size.\footnote{This approach to
the thermodynamic limit is conventional in the theory of
finite size scaling \cite{barber}.  There the critical
temperature has to be approached so that
$\xi/L$ is finite, where $\xi$ is the correlation length and
$L$ the linear system size.}
For the Ising model coupled to gravity 
$\xi_M \sim t^{-\nu} = t^{-3/d_H}$.
On the other hand, the linear size of the surface is 
$L \equiv A^{1/d_H} \leq l^{2/d_H}$, where we have used the behavior
of the scaling variable $z$.  Now requiring $\xi_M \leq L$
implies $t \geq l^{-2/3}$.  In this limit the second term
in the denominator of Eq.(~\ref{*38}) dominates, leading
to the area scaling
\begin{equation}
\label{*39}
\bar{A} \;\sim\; l^{4/3}.
\end{equation}
This together with Eq.~(\ref{*26}) yields Hausdorff dimension
\begin{equation}
\label{*395}
d_H \;=\; 4,
\end{equation}
which agrees with numerical simulations of the
Ising model on dynamical triangulations of spherical   
topology \cite{num1,num2,num3}. 

If we tune the deviation from the critical temperature more rapidly to
zero than $1/l^{2/3}$, Eq.~(\ref{*38}) gives us the ``classical'' area
scaling $\bar{A} \sim l^2$.  This implies $d_H = 6$, which coincides
with the result obtained in \cite{iikmns,kost2}.

\section{The continuum limit and Hausdorff dimension}

In the previous section we derived the result 
\begin{equation}
\label{*381}
\bar{A} \;=\; \frac{l^2 \; K_{1/3}(z)}
  {2\,z K_{4/3}(z) \; + \; 
    2^{2/3}\,t\,l^{2/3}\,z^{1/3} K_{2/3}(z) } \; ,
\end{equation}
where the scaling variable $z = u^{3/2}(\mu,t)l/\sqrt{2}$. 
We now extend our discussion of the nature of the continuum limit
in these models. Ultimately we must take the three different continuum limits:
\begin{enumerate}
\item $\mu \rightarrow \mu_c$ \qquad $(\bar{A} \rightarrow \infty)$
\item $\mu_b \rightarrow \mu_b^c$ \qquad $(l \rightarrow \infty)$ 
\item $t \rightarrow 0$ \qquad $(T \rightarrow T_c )$. 
\end{enumerate}
The first limit corresponds to simply letting the volume of the system
diverge.  The second limit corresponds to tuning the boundary
cosmological constant ($\mu_b$) to its critical value so that the
length of the boundary diverges along with the area. And the last
limit corresponds to adjusting the temperature so that the Ising spins
are critical.  If the mean area ($ \bar{A}$) diverges like
$\bar{A}\simeq l^{\nu}$, as we take $t \rightarrow 0$, then we have
$d_H= 3 \nu$, since the geodesic distance scales as $r \simeq
l^{1\over 3}$ for the Ising model.

The correct procedure, we claim, is to take the limits (1) and (2)
first. The situation is analogous to that of the Ising model in a
small residual external magnetic field. In that case one is interested
in the scaling behavior of the free energy as the scaling variable $x
= h/t^{\Delta}$ approaches zero. At any small but finite $h$, one
cannot take the limit $t \rightarrow 0$, as one then gets cross-over
to the critical behavior associated with the sink $h = \infty$ (i.e
the critical point associated with large $x$ behavior). Similarly here
we cannot take the limit $t \rightarrow 0$ at any finite $l$. In the
limit $l \rightarrow \infty$ first, we have $\nu=4/3$ and $d_H=4$.  As
mentioned, this approach to the continuum limit is also familiar in
the theory of finite-size scaling.  One has to take the continuum
limit in such a way that the matter correlation length is less than
the system size, otherwise the physics is dominated by boundary rather
than bulk effects.

The line with $y^{-1}=tl^{2/3}$ held constant may be viewed as analogous to 
a cross-over line. For $y$ small, with $t$ sufficiently small of
course to be near the critical point, we have the true 
critical behavior in which we are interested. For $y$ large the system
is dominated by boundary effects. There should thus be a cross-over
for $y$ in the neighborhood of $1$. It is difficult to say exactly how rapid
this cross-over is. But eventually one might see the emergence of 
Hausdorff dimension 6 scaling for $t$ sufficiently small at finite
size $l$.  It may also be that this regime is never 
reached. There is no evidence for such a regime in the present numerical
simulations but it should be looked for more carefully.

\section{Discussion}

In this paper we examine how the 
intrinsic dimensionality of two-dimensional gravity depends
on its coupling to unitary $(m+1,m)$ conformal field theories
for the case of pure gravity $(m=2)$ and the Ising model $(m=3)$.
In such theories one must take the 
thermodynamic limit simultaneously with the approach
to the critical point of the matter.
In particular we are interested in the scaling of the mean
area of a disk with its boundary length.
We have shown that this scaling depends on the
precise approach to the infinite volume critical point. 
In the continuum
limit one has to consider how the correlation length of the matter fields
diverges in relation to the gravitational correlation length.
We find that for the Ising model perturbed by a thermal operator
the mean area does not scale according to the classical result
$A \sim l^2$, but rather with an anomalous exponent that depends on the
matter. Combined with the scaling of geodesic distance with boundary
length, following from $c < 1$ string field theory arguments, 
we find that the Hausdorff dimension of surfaces in 
$2d${--}gravity coupled to the Ising is 4. 
Tuning to the critical temperature before taking the infinite volume
limit yields $d_H=6$.

Our results suggest some further avenues of investigation.  A critical
identification in this paper is that between geodesic distance and
proper time in the string field theory -- this may simply be wrong when
matter is coupled.  For a recent discussion of this point 
and a determination of the Hausdorff dimension 
for $c=-2$ matter see \cite{aaijkwy}.
The behavior of minimal models (coupled to
gravity) on the disk is intriguing from several points of view,
particularly if one adds a boundary magnetic field \cite{taylor2}.
Our predicted scaling behavior of the mean area with boundary length
on the disk should certainly be checked in numerical simulations.

Although we have restricted our analysis to the case of the pure
gravity and the Ising model coupled to gravity these issues remain
relevant to the generic $(p,q)$ minimal model coupled to gravity.  There,
however, the space of relevant operators is much larger and more
intricate and this will lead to a subtle interplay of the different length
scales introduced by various perturbations. As we have shown, though,
the situation is complicated enough for the Ising model, and a better
understanding of even this simple case would greatly clarify the
manner in which geometrical quantities behave in reparametrization
invariant theories like quantum gravity.
  
\section*{Acknowledgements}

We would like to thank Jan Ambj{\o}rn, Kostas Anagnostopoulos and
Nobuyuki Ishibashi for detailed discussions of non-critical string
field theory and issues related to the fractal dimensionality of
two-dimensional gravity.  We have benefited as well from useful
conversations with Simon Catterall, Marco Falcioni and Fran\c{c}ois David.
This research was supported by the Department of Energy, USA, under 
contract No.DE-FG02-8540237 and by research funds from Syracuse University.

\flushleft

\end{document}